\documentclass[iop]{emulateapj} 
\usepackage{natbib,threeparttable,apjfonts,graphicx,booktabs}

\shortauthors{Castro et al.}
\slugcomment{Accepted for publication in ApJ}

\begin{document} 
\title{THE IMPACT OF EFFICIENT PARTICLE ACCELERATION ON THE EVOLUTION OF SUPERNOVA REMNANTS IN THE SEDOV-TAYLOR PHASE}
\author{Daniel Castro\altaffilmark{1,2}, Patrick Slane\altaffilmark{1}, Daniel J. Patnaude\altaffilmark{1}, and Donald C. Ellison\altaffilmark{3}}

\altaffiltext{1}{Harvard-Smithsonian Center for Astrophysics, 60 Garden Street, Cambridge, MA 02138, USA}
\altaffiltext{2}{Departamento de F\'isica, Universidad Sim\'{o}n Bol{\'\i}var, Valle de Sartenejas, Apdo. 89000, Caracas 1080A, Venezuela}
\altaffiltext{3}{Department of Physics, North Carolina State University, Box 8202, Raleigh, NC 27695, USA}

\begin{abstract}
We investigate the effects of the efficient production of cosmic rays on the evolution of supernova remnants (SNRs) in the adiabatic Sedov-Taylor phase. We model the SNR by coupling the hydrodynamic evolution with nonlinear diffusive shock acceleration (DSA), and track self-consistently the ionization state of the shock-heated plasma. Using a plasma emissivity code and the results of the model, we predict the thermal X-ray emission and combine it with the non-thermal component in order to obtain the complete spectrum in this energy range. Hence, we  study how the interpretation of thermal X-ray observations is affected by the efficiency of the DSA process, and find that, compared to test particle cases, the efficient DSA example yields a smaller shock radius and speed, a larger compression ratio, and lower intensity X-ray thermal emission. We also find that a model where the shock is not assumed to produce CRs can fit the X-ray observational properties of an example with efficient particle acceleration, with a different set of input parameters, and in particular a much lower explosion energy. Additionally, we model the broadband non-thermal emission, and investigate what signatures result from the acceleration of particles. 
\end{abstract}


\section{Introduction}

A supernova remnant (SNR) is the result of the interaction between supernova (SN) ejecta and the surrounding interstellar medium (ISM). As the SN ejecta expands it sweeps up the surrounding material, and forms a forward shock wave (FS) that compresses the ISM heating it to X-ray emitting temperatures. The deceleration of the SN ejecta also produces a reverse shock (RS) that travels inward from the contact interface, or contact discontinuity (CD), heating the ejected material. The evolution of SNRs and their emission characteristics can be divided into several different phases \citep{Chevalier1977}. Initially the expansion and thermal X-ray emission properties of an SNR are dominated by the ejecta. Once the majority of the ejecta energy has been transferred to the surrounding medium, the SNR enters the Sedov-Taylor phase. During this stage the hydrodynamic evolution can be approximated by the self-similar adiabatic expansion of a shock wave originating at a point explosion and propagating through an uniform medium \citep{Sedov1959,Taylor1950}. For SNRs in this stage, the Sedov-Taylor similarity solution for the expansion of the blast wave, together with the Rankine-Hugoniot relations \citep{Rankine1870,Hugoniot1889}, allow parameters of the SN and ambient medium to be estimated from thermal X-ray emission observations \citep{Hamilton1983}. 
%A supernova remnant (SNR) is the result of the interaction between supernova (SN) ejecta, which carries most of the SN energy, and the surrounding interstellar medium (ISM). As the SN ejecta expands it sweeps up the surrounding material, and forms a forward shock wave (FS) that compresses the ISM heating it to X-ray emitting temperatures. The deceleration of the SN ejecta also produces a reverse shock (RS) that travels inward from the contact interface, or contact discontinuity (CD), heating the ejected material. The evolution of SNRs and their emission characteristics can be divided into several different phases \citep{Chevalier1977}. Initially the expansion and thermal X-ray emission properties of an SNR are dominated by the ejecta. Once the majority of the ejecta energy has been transferred to the surrounding medium, the SNR enters the Sedov-Taylor phase. During this stage the hydrodynamic evolution can be approximated by the self-similar adiabatic expansion of a shock wave originating at a point explosion and propagating through an uniform medium \citep{Sedov1959,Taylor1950}. For SNRs in this stage, the Sedov-Taylor similarity solution for the expansion of the blast wave, together with the Rankine-Hugoniot relations \citep{Rankine1870,Hugoniot1889}, allow parameters of the SN and ambient medium to be estimated from thermal X-ray emission observations \citep{Hamilton1983}. 

In addition to heating the surrounding medium, SNR shocks are thought to accelerate ambient particles to cosmic ray (CR) energies, through diffusive shock acceleration (DSA). Non-thermal X-ray emission has been detected from young shell-type SNRs, including SN 1006 \citep{Koyama1995,Reynolds1998}, RX J1713.7-3946 \citep{Koyama1997,Slane1999}, and Vela Jr. \citep{Aschenbach1998,Slane2001}. These X-rays are believed to be synchrotron radiation from electrons accelerated to TeV energies at the SNR shock. Additionally, if a significant fraction of the SN energy is placed in relativistic particles the hydrodynamic evolution of the SNR will be modified, and some evidence of this has been observed. X-ray observations of the relative positions of the FS and RS, and contact discontinuity (CD), in Tycho's SNR, and SN 1006 point to the shock compression ratios being modified by the acceleration of cosmic-ray ions \citep{Warren2005,Cassam2008}. Also, the postshock plasma temperatures observed in SNRs 1E 0102.2-7219 and RCW 86, are lower than expected for their measured shock velocities, and hence also indicate that particle acceleration at their shocks is efficient \citep{Hughes2000,Helder2009}.

The evidence of efficient acceleration of cosmic rays at SNR shocks suggests that the DSA process must be incorporated into the analysis of the evolution of remnants, and some early work has outlined the impact of particle acceleration on observations of SNRs in the adiabatic stage \citep{Heavens1984,Dorfi1993}. Here, we investigate the effects of efficient DSA on the evolution of SNRs in the Sedov-Taylor phase, and how it impacts the understanding of X-ray observations of remnants. To this end we use a hydrodynamic model coupled with nonlinear DSA, and with a self-consistent treatment of the non-equilibrium ionization (NEI) structure of the shock-heated plasma. 

In Section 3.2, we outline the Sedov-Taylor solution for the adiabatic expansion of SNRs, and the Rankine-Hugoniot equations for the propagation of shock waves. Additionally, we show how the relations between the physical parameters are modified when a fraction of the SN energy that goes into compressing the ambient medium changes, and the equation of state of the plasma is softened due to the acceleration of particles at the shock. We describe our model in more detail in Section 3.3, and present simulation examples of SNRs with both efficient and inefficient particle acceleration and discuss these results in Section 3.4.

\section{The Sedov-Taylor Phase and Particle Acceleration}

The Sedov-Taylor self-similar solution describes the expansion of a blast wave that originates in a point explosion and propagates in a homogeneous medium \citep{Sedov1959, Taylor1950}. SNR shocks are believed to relax towards this description once the mass of swept-up ISM is larger than that of the SN ejecta. In this section we present a simple analytical treatment that connects the SN explosion energy, the age, and the ISM density with parameters that can be obtained from observing thermal X-ray emission from the shock-heated plasma in SNRs. 

For an SNR that transfers energy $E$ into the heating of the surrounding plasma, with ISM pre-shock density $\rho_0$, the expansion of the SNR at age $t$ is described by the expression

\begin{equation}\label{eq:3.1}
R_S = \left[\frac{\alpha(\gamma)E t^2}{\rho_0}\right]^{1/5}\, ,
\end{equation}

\noindent where $\alpha$ is a function of the specific heat ratio of the gas $\gamma$, and is calculated numerically to solve the self-similar problem \citep[see Figure 75]{Sedov1959}. However, as one expects a fraction $\theta$ of the total SN explosion energy, $E_0$, to be placed in CRs, then we can define $\beta=[\alpha(1-\theta)]^{1/5}$, and hence

\begin{equation}\label{eq:3.2}
R_S = \beta(\gamma,\theta)\left[\frac{E_0 t^2}{\rho_0}\right]^{1/5}\, .
\end{equation}

\noindent From this result we can also derive the shock velocity to be $v_S = 2 R_S /(5 t)$.

The application of conservation laws across the shock front yield the Rankine-Hugoniot relations, which connect post-shock parameters, like temperature $T$ and density $\rho_1$, to the pre-shock conditions \citep[and references therein]{McKee1987}. For strong shocks, where the ratio of the shock velocity to the upstream signal velocity, or Mach number $M$, is much greater than 1, the compression ratio is given by 

\begin{equation}\label{eq:3.3}
\frac{\rho_1}{\rho_0}=\frac{\gamma+1}{\gamma-1}\, .
\end{equation}

\noindent Furthermore, the pressure $P_S$ of the post-shock material in this case is 

\begin{equation}\label{eq:3.4}
P_S=\frac{2\rho_0 v_S^2}{(\gamma+1)}\, ,
\end{equation}

\noindent where we can replace the pressure using the ideal gas law, $P=\rho k T/(\mu m_{\text{H}})$, where $T$ is the temperature of the gas, and $k$ is the Boltzmann constant, resulting in

\begin{equation}\label{eq:3.5}
kT_S =\frac{2(\gamma-1)} {(\gamma+1)^2}\,\mu m_{\text{H}} v_S^2\, .
\end{equation}

\noindent In this expression $\mu$ is the mean atomic weight, and $m_{\text{H}}$ is the mass of the hydrogen atom. In the case of Sedov-Taylor SNRs we can express the shock velocity $v_S$ in terms of shock radius and age, and hence

\begin{equation}\label{eq:3.6}
kT_S =\frac{8(\gamma-1)} {25(\gamma+1)^2} \,\mu m_{\text{H}} \left(\frac{R_S}{t}\right)^2 \, .
\end{equation}

In order to express the emission measure of the plasma $\xi = n_e n_{\text{H}} V$, where $n_e$ and $n_{\text{H}}$ are the post-shock electron and hydrogen number densities respectively, in terms of the pre-shock conditions we need to consider the geometry of this simplified problem. The SNR shock in this basic picture is expanding spherically and compressing the ISM into a shell. From mass conservation considerations, the shell volume, $V$, must follow the relation $\rho_1 V = \rho_0 4\pi R_S^3/3$, and using the compression ratio, as in Equation \ref{eq:3.3}, 
 
\begin{equation}\label{eq:3.7}
V =\frac{(\gamma-1)} {(\gamma+1)} \, \frac{4}{3}\pi R_S^3 \, .
\end{equation}

\noindent Using Equations \ref{eq:3.3} and \ref{eq:3.7}, we get the volume emission measure:
 
\begin{equation}\label{eq:3.8}
\xi =\frac{(\gamma+1)} {(\gamma-1)} \left(\frac{n_e}{n_H}\right) n_0^2 \,\frac{4}{3}\pi R_S^3 \, ,
\end{equation}

\noindent where $n_0$ is the pre-shock proton number density. 

If the shock is not assumed to accelerate particles to CR energies then Equations \ref{eq:3.2}, \ref{eq:3.6}, and \ref{eq:3.8}, can be used to estimate $E_0$, $\rho_0$, and $t$, from the values of radius, plasma temperature, and emission measure, obtained from observations. In that case, when the fraction of the SN explosion energy that is placed in relativistic particles is $\theta=0$, and the adiabatic index of the plasma is $\gamma=5/3$, and also assuming a mean atomic weight, $\mu=0.6$, and a ratio of electron to hydrogen number density $n_e/n_{\text{H}}=1.23$, we obtain the following expressions:

\begin{equation}\label{eq:3.9}
n_0=40.64 \left(\frac{\xi}{10^{60} \,\text{cm}^{-3}}\right)^{1/2} \left(\frac{R_S}{ 1 \,\text{pc}}\right)^{-3/2}\,\, \text{cm}^{-3}\, ,
\end{equation}

\begin{equation}\label{eq:3.10}
t=423.9 \left(\frac{kT_S}{1\,\text{keV}}\right)^{-1/2} \left(\frac{R_S}{ 1 \,\text{pc}}\right)\,\,\text{yr}\, ,
\end{equation}

\begin{equation}\label{eq:3.11}
E_0=0.075 \left(\frac{\xi}{10^{60} \,\text{cm}^{-3}}\right)^{1/2} \left(\frac{kT_S}{1\,\text{keV}}\right)\left(\frac{R_S}{ 1 \,\text{pc}}\right)^{3/2}\times 10^{51}\text{ergs}\, .
\end{equation}

Cosmic ray acceleration at the shock modifies this analysis, since some of the explosion energy goes into relativistic particles, and hence $\theta>0$. Efficient particle acceleration also makes this a two-fluid problem, with a thermal plasma of adiabatic index $\gamma=5/3$, and a relativistic gas of CRs with $\gamma=4/3$. \citet{Chevalier1983} uses enthalpy arguments to address this problem analytically, and defines an effective specific heat ratio,
\begin{equation}\label{eq:3.12}
\gamma_S = \frac{5+w}{3(1+w)}\, ,
\end{equation}

\noindent where $w=p_{cr}/p_s$ is the fraction of the post-shock pressure in relativistic particles, which is dependent on the fraction of the total energy going into particle acceleration. Additionally, a fraction of the high energy particles accelerated at the shock is expected to escape from the shock, further softening the equation of state in the shocked plasma \citep{Berezhko1999}.

\begin{figure}[h!]
\begin{center}
\includegraphics[width=\columnwidth]{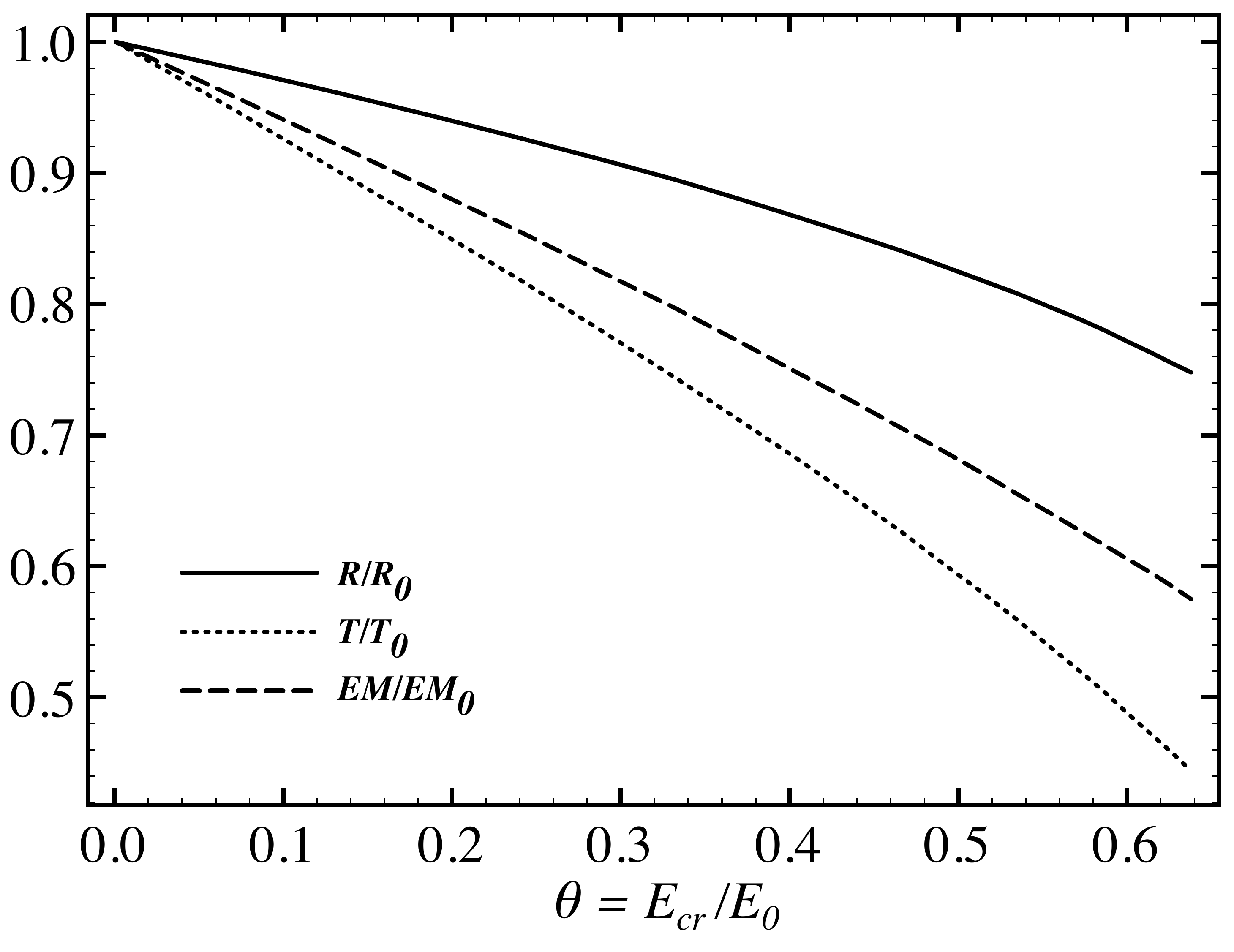}
\caption{Profile of the radius (solid), temperature (dotted), and emission measure (dashed) as functions of the energy placed in CRs, relative to the values of these parameters for the case where no efficient acceleration is taking place ($\theta=0$ and $\gamma=5/3$).}
\label{fig:ratios}
\end{center}
\end{figure}

The softening of the equation of state of the plasma, and lower energy available for plasma heating, result in higher compression ratios and lower post-shock temperatures. Figure 1 illustrates how the radius, the temperature and the emission measure, calculated using Equations \ref{eq:3.2}, \ref{eq:3.6}, and \ref{eq:3.8}, change as a function of the fraction of SN explosion energy placed into CRs. The variation of the effective adiabatic index of the post-shock plasma with $\theta$ is determined using the results from \citet{Chevalier1983}. 

\begin{figure}[h!]
\begin{center}
\includegraphics[width=\columnwidth]{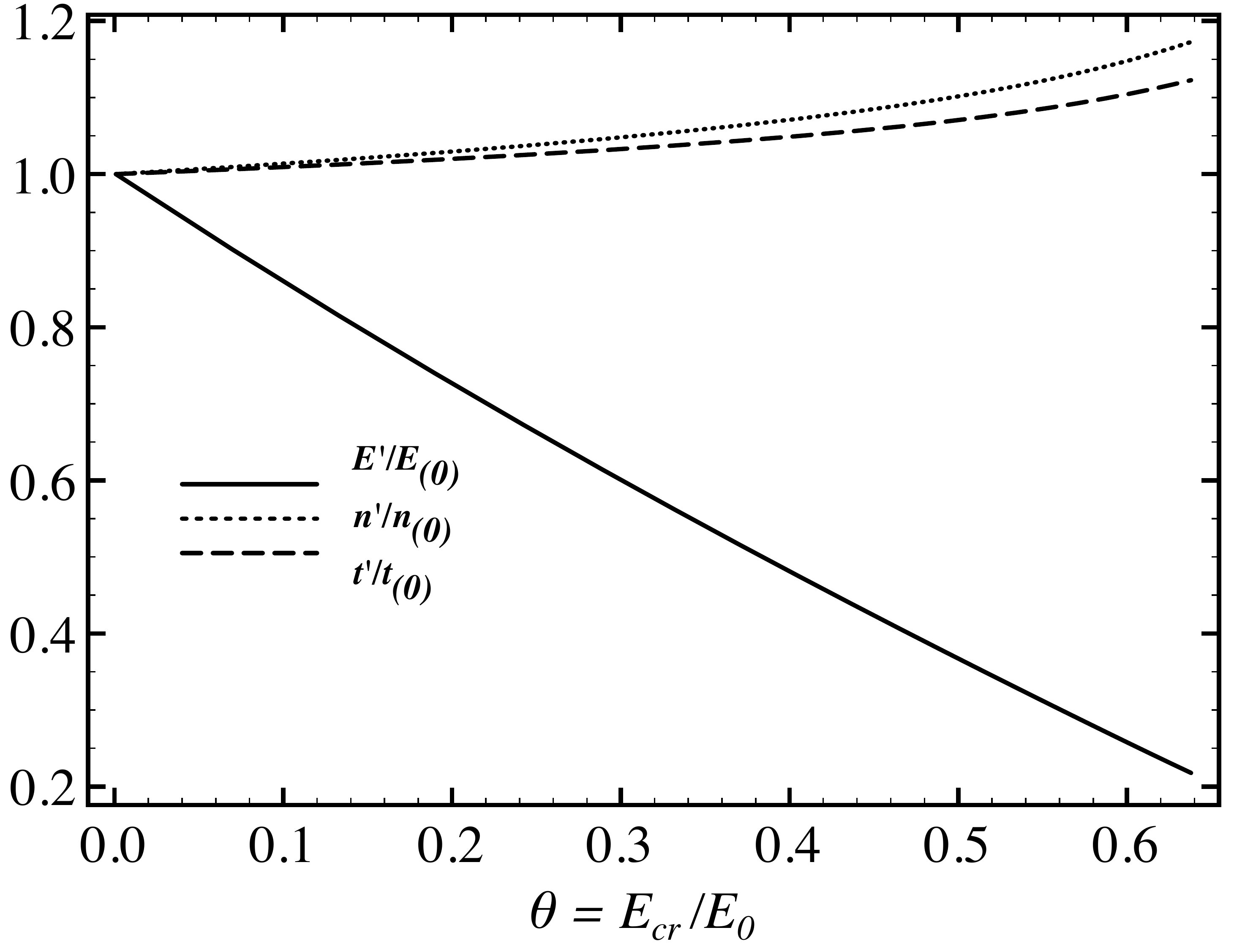}
\caption{Approximate trends of the inferred SN explosion energy $E_0$, age $t$, and ambient proton number density $n_{\text{0}}$, as a function of the fraction of the SN explosion energy placed in relativistic particles, assuming no efficient acceleration is taking place ($\theta=0$ and $\gamma=5/3$). }
\label{fig:ratios2}
\end{center}
\end{figure}

 In Figure \ref{fig:ratios2}, we show how the retrieved values of explosion energy, ambient density, and age of the SNR, obtained using Equations \ref{eq:3.9}-\ref{eq:3.11}, which assume that no CR acceleration has taken place, deviate from the input parameters used in Equations \ref{eq:3.2}, \ref{eq:3.6}, and \ref{eq:3.8}, if the shock places some of its energy into the acceleration of CRs, i.e., $\theta>0$. This simple approach is a useful illustration of the approximate modified trends of the Sedov-Taylor analysis of the evolution of SNRs where efficient particle acceleration at the FS is taking place. However it does not include several important factors in the evolution of SNRs such as the free expansion phase, when the dynamics of the remnant are dominated by the expansion of the SN ejecta, and the temporal variation of the effective adiabatic index of the plasma. Nonetheless, the plot identifies the expected trends for parameters derived under the assumption that no CR acceleration has occurred.

\newpage

 \section{The CR-hydro model with X-ray emission}
 
The CR-hydro model is a hydrodynamic simulation of an SNR that incorporates particle acceleration at the shock, and determines the ionization state of the shocked plasma at every time step. The X-ray emission profile is then calculated, using the NEI information for the thermal component, and the electron spectrum for the simulated synchrotron emission. This model is described in more detail in \citet{Ellison2007}, \citet{Patnaude2009,Patnaude2010}, and references therein.

CR-hydro calculates the hydrodynamic evolution of an SNR with a spherically symmetric model described in \citet{Ellison2007}, and references therein. A semianalytical model of nonlinear DSA by \citet{Amato2005} and \citet{Blasi2005}, is coupled to the hydrodynamic calculation. Given an input DSA efficiency of the shock $\varepsilon_{\text{DSA}}$, the model determines the appropriate injection parameter ($\xi$ in Equation [25] in \citealt{Blasi2005}) as the system evolves. The injection parameter is then used to solve the DSA problem, together with the shock information and ambient density, temperature, and magnetic field. Hence, the energy loss from escaping accelerated particles is removed from the shocked plasma, and the adiabatic index of the shocked gas, $\gamma_{\text{sk}}$, is determined from the particle distribution function, which includes the appropriate ratio of relativistic and non-relativistic particles. This adiabatic index is then used in the hydrodynamic calculation, coupling DSA to the shock evolution. 

\begin{figure}[h!]
\begin{center}
\includegraphics[width=0.85\columnwidth]{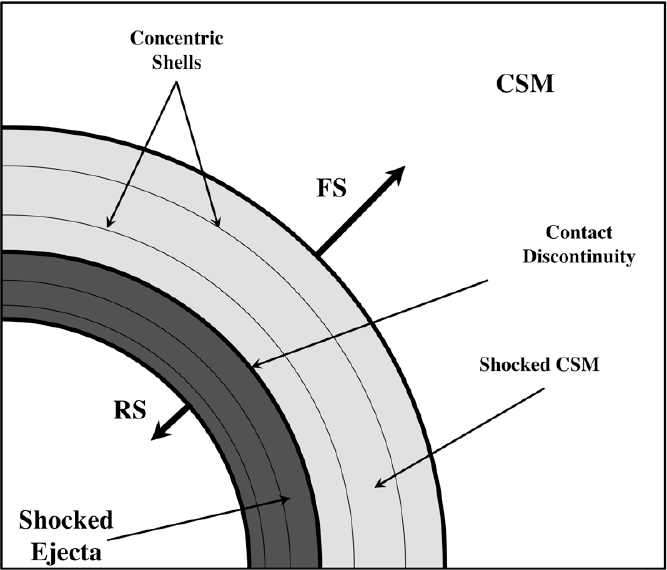}
\caption{Diagram of the SNR model used in CR-hydro, which shows the positions of the FS, RS, and CD. The model is divided in concentric spherical shells of shocked material overtaken by the FS, which evolve through time.}
\label{fig:diagram}
\end{center}
\end{figure}

CR-hydro tracks the time-dependent hydrodynamics of the system in Lagrangian mass coordinates, subdividing it in a series of concentric shells of shocked material. As the mass of each shell remains constant, the volume of the shell as the system evolves is adjusted to account for variations of the pressure. Both the RS and FS are included in the model, and shocked ejecta and shocked ISM are separated by a contact interface. As the shocks overtake new material, new shells are added to the system. A schematic view of the geometry of the model is shown in Figure \ref{fig:diagram}.

At each time step, the NEI state of the shocked plasma, between the FS and CD, is calculated self-consistently, using the ionization structure, free electron number density, and electron temperature. An updated version of the \citet{Raymond1977} plasma emissivity code is then used to obtain the thermal X-ray emission from the system \citep{Brickhouse1995}, where the resulting NEI fractions of heavy elements, and plasma temperature and density, are used as input parameters. \citet{Patnaude2009}, and \citet{Patnaude2010}, provide a detailed description of this NEI model. 

The resulting electron spectrum from the acceleration process is used to calculate the synchrotron radiation from the system, as well as the IC emission assuming standard seed photons from the cosmic microwave background. Additionally, the relativistic Bremsstrahlung photon spectrum is obtained using the spectral distribution of accelerated electrons. The model also calculates the pion decay emission spectrum, resulting from proton-proton collisions, based on the population of protons accelerated to relativistic speeds at the FS, and using the parameterization described in \citet{Kamae2006}. 

 An important factor determining the shape of the synchrotron spectrum is the nature of the magnetic field. Because the compression of the fluid is altered by DSA, the compression of the magnetic field is also expected to be affected by the efficiency of this process. The ad hoc compression model of the magnetic field used in this simulation is described in \citet{Ellison2005}. This treatment uses the far upstream magnetic field to calculate the Alfv\'en heating rate in the shock precursor, which in turn results in changes in the subshock compression and the acceleration efficiency of the system. Magnetic field amplification (MFA) at the FS is now believed to be an important effect in young SNR shocks (e.g., \citealt{Vink2003,Uchiyama2007}), and is included in the model as an ad hoc magnification factor, set explicitly as an input parameter, $B_{\text{amp}}$. The compressed field immediately behind the shock is increased by a factor $B_{\text{amp}}$. The resulting compressed field is used to calculate the synchrotron emission from the accelerated electron population.

We study a set of models where we explore different values of the ejecta kinetic energy from the SN explosion $E_{\text{SN}}$, and the acceleration efficiency, $\varepsilon_{\text{DSA}}$. In all examples in this work the model is initialized using input parameters similar to those of Type Ia supernovae, i.e, mass of the ejecta $M_{\text{ej}}=1.4 M_{\odot}$, exponential density profile of the ejecta \citep{Dwarkadas2000}, and kinetic energy in ejecta from the SN explosion $E_0=10^{51}$ erg. We assume a uniform ambient proton density $n_{\text{p}}$, and magnetic field $B_0=5\mu\text{G}$. Additionally, we select explicitly the value of the relativistic electron to relativistic proton ratio $K_{{ep}}=2\%$, in line with estimates inferred from direct measurements of cosmic ray spectra at Earth. This parameter determines how thermal and non-thermal X-ray emission scale relative to each other, as well as whether the $\gamma$-ray emission is dominated by leptonic or hadronic emission processes. Further discussion of the input parameters required for the CR-hydro model is provided in \citet{Ellison2007} and \citet{Patnaude2009}. In our next work, we intend to apply the model presented here to real cases of SNRs, for which it will be crucial to explore the parameter space extensively.

The thermal X-ray emission model obtained from the plasma emissivity code is combined with the non-thermal emission in this range to obtain the complete X-ray emission profile. In order to simulate actual observations of SNRs in X-rays, the resulting spectrum model is then folded through the instrument response of the Advanced CCD Imaging Spectrometer (ACIS) on board the \emph{Chandra X-ray Observatory}. The distance to the SNR is fixed at 5 kpc, and the Galactic absorbing column density is set to be $N_{\text{H}}=5\times10^{21} \text{atoms cm}^{Ð2}$. No background contribution is included in the model or the subsequent analysis.

\section{Results and Discussion}

In order to investigate how SNRs in the Sedov-Taylor phase are affected by efficient particle acceleration, we study a set of examples where the model is allowed to evolve to ages where the swept up ISM mass is much greater than the ejecta mass. First, we explore the test particle (TP) case, where the particle acceleration process is not efficient; then, using the same set of input parameters, we study the case were DSA is efficient; and finally, we consider how the efficient case can be interpreted, and compare its results to those of test particles models.

\subsection{Test particle and efficient acceleration cases}

We consider the TP case by setting the DSA efficiency of the model to $\varepsilon_{\text{DSA}}=0.1\%$, and the ambient proton density to $n_{\text{p}}=1$ cm$^{-3}$. The model is allowed to evolve to $t_{\text{SNR}}=10,000$ yr, and at the end of the simulation we obtain the following parameters: FS radius $R_{\text{FS}}=12.6$ pc; FS speed $V_{\text{FS}}=510 $ km s$^{-1}$; overall FS compression ratio $R_{\text{tot}}=4.0$; CD radius $R_{\text{CD}}=8.5$ pc; and a swept up mass $M_{\text{sw}}=280 M_{\odot}$. The fraction of the total explosion energy placed into cosmic rays in this example is $\theta=1.6\%$.

The expected value of the radius of the FS for this set of parameters can be derived from Equation \ref{eq:3.1}, which yields $R_{\text{S}}=12.5$ pc. The small deviation from this value obtained from the CR-hydro model ($>$1\%) can be attributed to the fact that the simple form of the Sedov-Taylor solution that results in Equation \ref{eq:3.1}, does not take into consideration the initial phase of the SNR, when the SN ejecta freely expands into the ISM. 

For the efficient acceleration case we assume that 40\% of the shock ram kinetic energy flux is placed in superthermal particles, i.e., $\varepsilon_{\text{DSA}}=40\%$. At an age of 10,000 yr, and for $n_{\text{p}}=1$ cm$^{-3}$, the output parameters for this example are: FS radius $R_{\text{FS}}=12.1$ pc; FS speed $V_{\text{FS}}=480 $ km s$^{-1}$; overall FS compression ratio $R_{\text{tot}}=4.9$; CD radius $R_{\text{FS}}=8.6$ pc; and a swept up mass $M_{\text{sw}}=250 M_{\odot}$. In this case 28\% of the $10^{51}$ erg SN explosion energy  goes into the acceleration of particles at the shock.

\begin{figure}[h!]
%\begin{raggedleft}
\begin{center}
%\includegraphics[width=0.45\columnwidth]{f4a.pdf}\\
%\vspace{-0.1cm}
%\includegraphics[width=0.45\columnwidth]{f4b.pdf}\\
%\vspace{-0.1cm}
%\includegraphics[width=0.45\columnwidth]{f4c.pdf}\\
%\vspace{-0.1cm}
%\includegraphics[width=0.45\columnwidth]{f4d.pdf}\\
\includegraphics[width=\columnwidth]{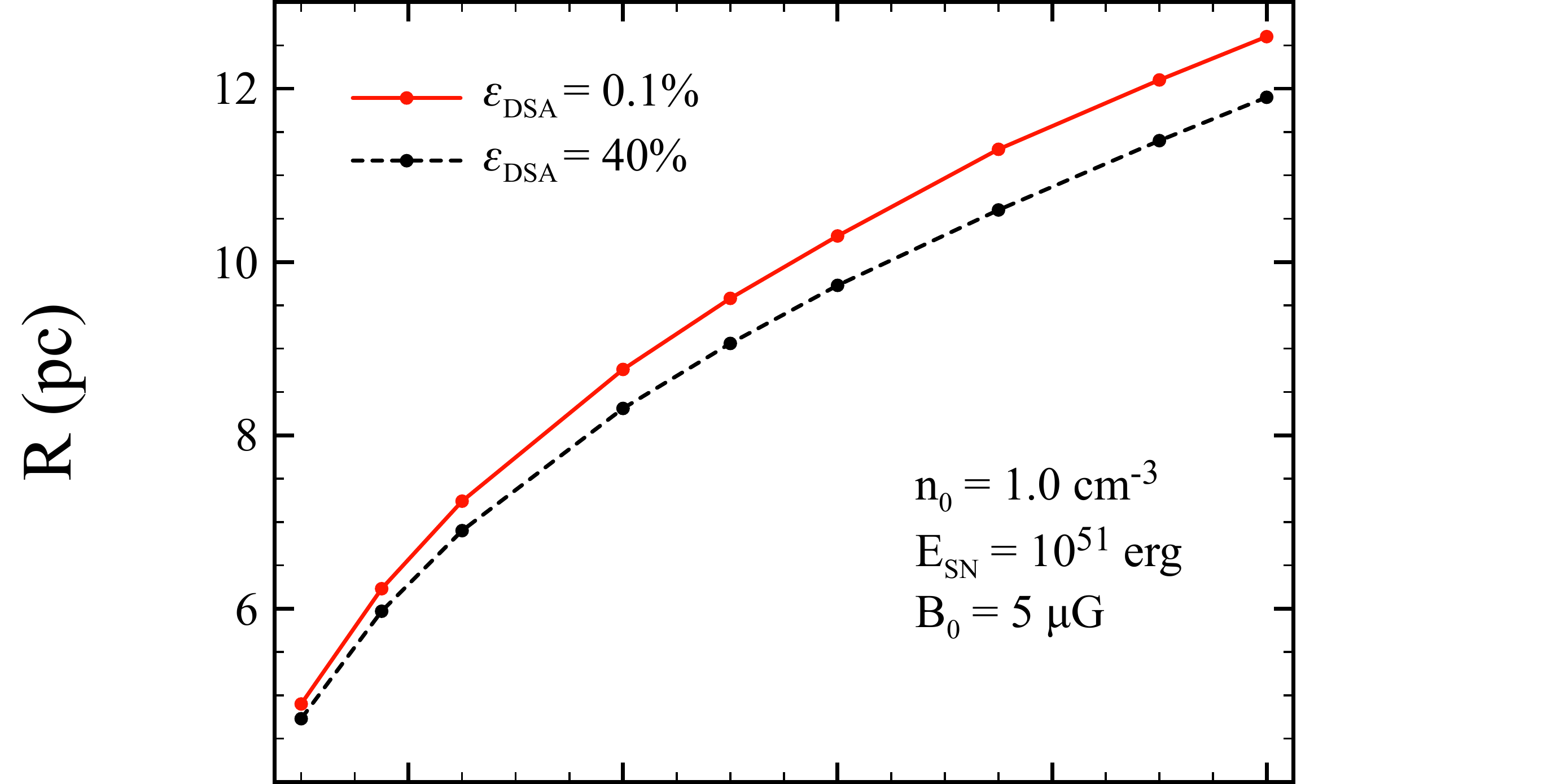}\\
\vspace{-0.1cm}
\includegraphics[width=\columnwidth]{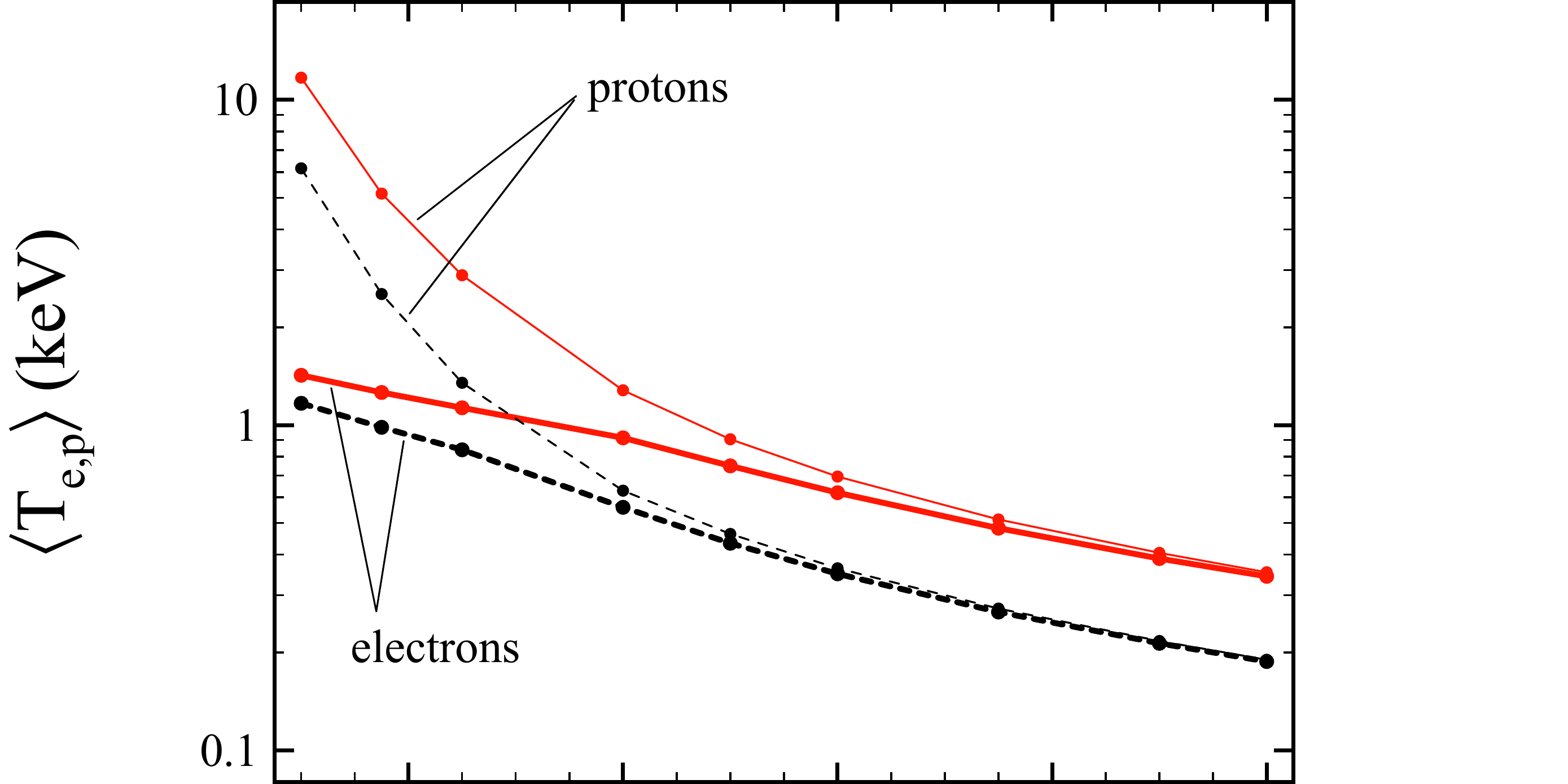}\\
\vspace{-0.1cm}
\includegraphics[width=\columnwidth]{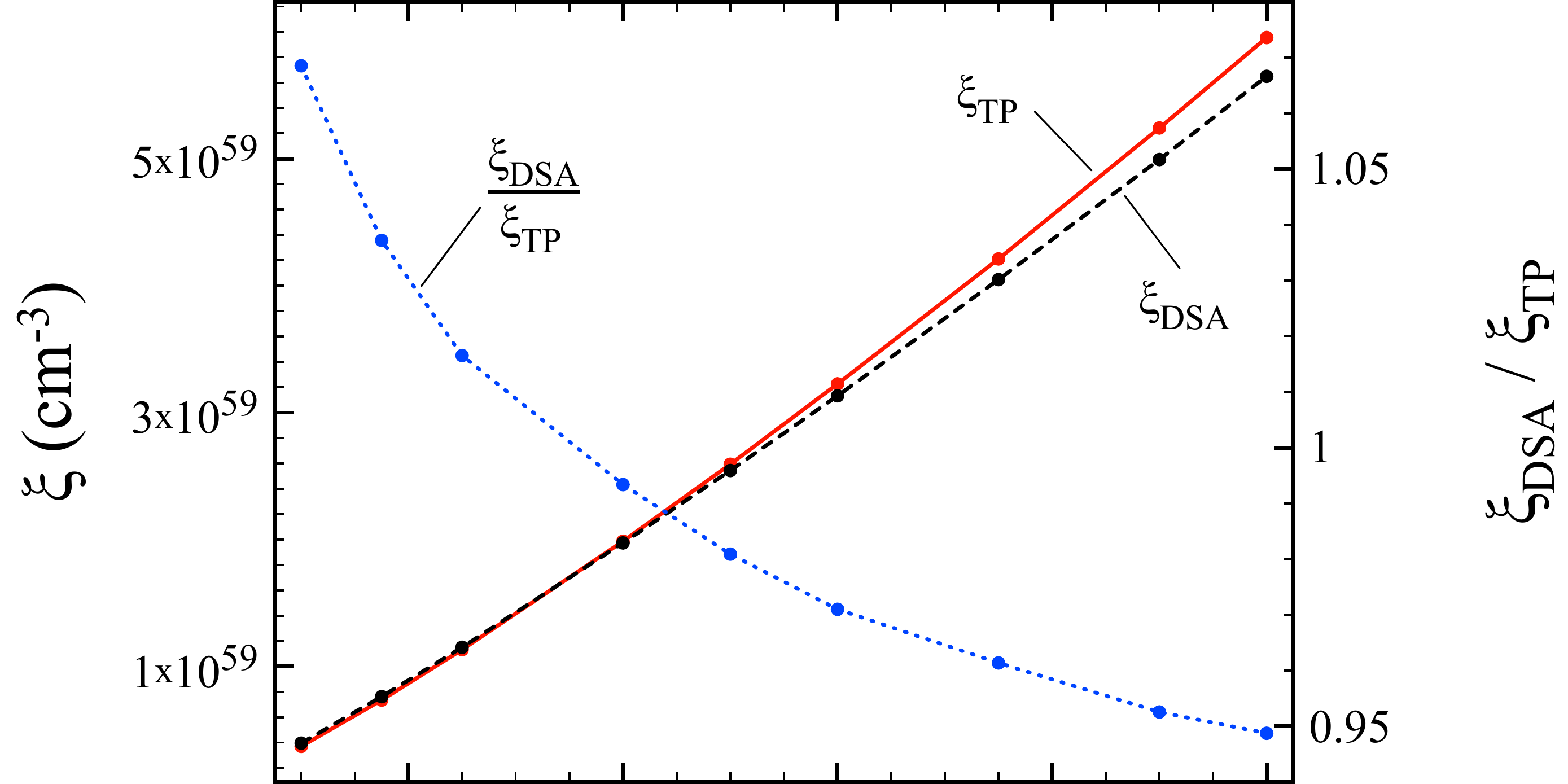}\\
\vspace{-0.1cm}
\includegraphics[width=\columnwidth]{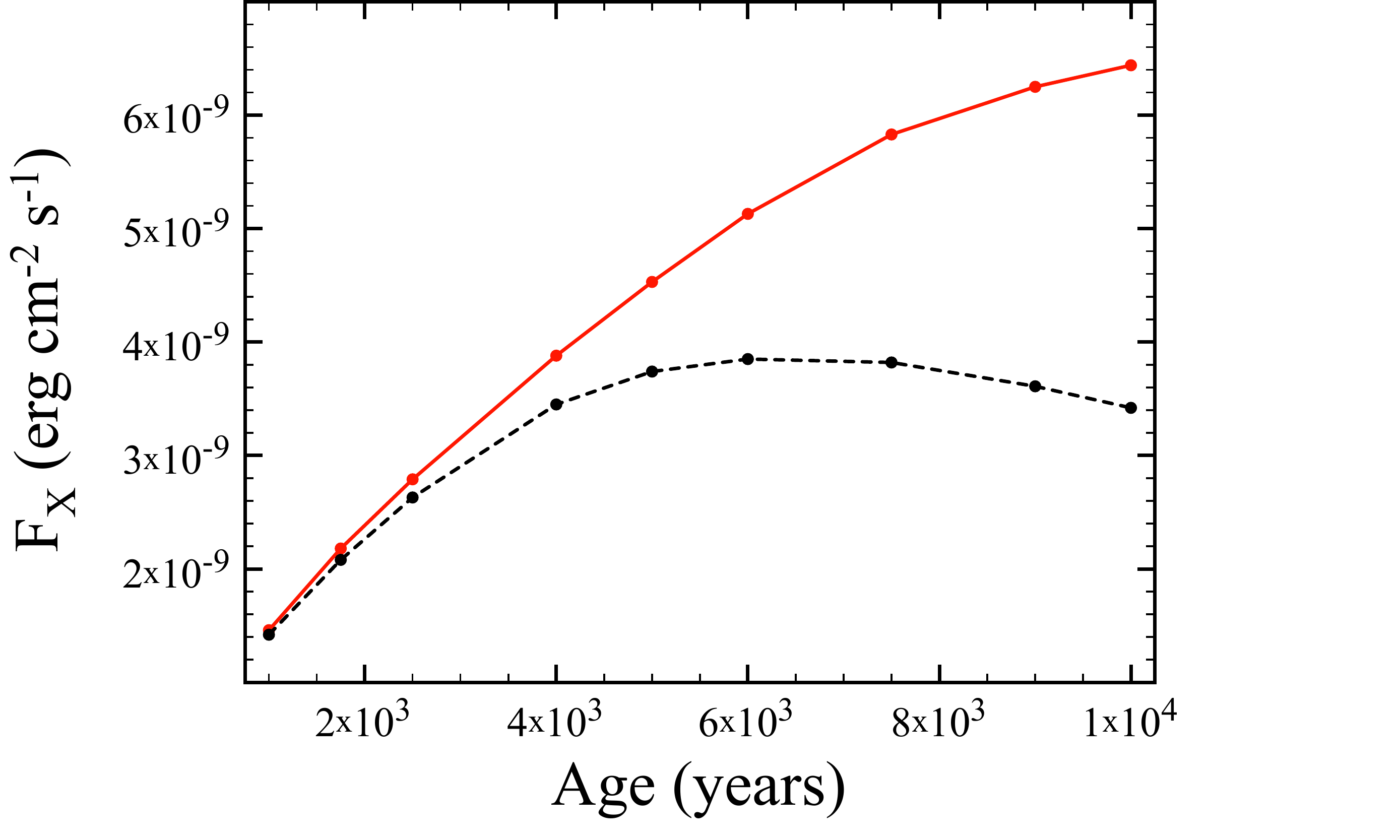}\\
\caption{The top panel shows the variation of the radius of the forward shock as a function of remnant age. As in all panels, the red solid line represents the TP model, and the efficient acceleration case with $\varepsilon_{\text{DSA}}=40\%$ is shown as a black dashed line. The  emission measure weighted mean temperatures of protons and electrons are presented in the second  panel. The third panel shows the evolution of the total volume emission measure of the shocked ISM, as well as the ratio of this parameter between the efficient and TP cases, presented as a blue dotted line (right axis). The bottom panel presents the thermal X-ray flux in the energy range 0.3-5.0 keV. The SNR models shown have input parameters $E_0=10^{51}$ erg, and $n_{\text{p}}=1.0$ cm$^{-3}$.}
\label{fig:evol}
%\end{raggedleft}
\end{center}
\end{figure}

The thermal X-ray emission is controlled in a complex way by the efficiency of the process of cosmic ray production. To understand this situation in more detail, the temporal evolution of the radius, proton and electron temperatures, volume emission measure, and thermal X-ray flux (in the range 0.3-5.0 keV), for the TP and DSA efficient cases are presented in Figure \ref{fig:evol}. For the input parameters studied, the forward shock radius (top panel) and particle temperatures (second panel) of the TP case are consistently larger than those of the efficient acceleration case. While the emission measure is expected to broadly scale as the compression ratio times swept up mass, in a complex hydrodynamic model such as the one presented here, the value of the total emission measure will also depend on how the shocked gas evolves after the initial compression, and on how the compression ratio varies with time. \citet{Ellison2007}, and \citet{Patnaude2009,Patnaude2010} find that for ages $t\lesssim10^3$ years, the volume emission measure increases with DSA efficiency. The third panel in Figure \ref{fig:evol} shows that, for the parameters used, the volume emission measure of the DSA efficient case is indeed larger than that of the TP case at $1,000$ years, yet the ratio $\epsilon_{\text{DSA}}/\epsilon_{\text{TP}}$ decreases with age, and at approximately $4,000$ years it becomes smaller than 1. Figure \ref{fig:evol} (bottom panel) shows that the thermal X-ray flux from the TP case becomes increasingly more intense with age than that of the DSA efficient case, which is the combined result of the temporal evolution of the emission measures, electron temperatures, and ionization conditions. 

\begin{figure}[h!]
\begin{center}
\includegraphics[width=\columnwidth]{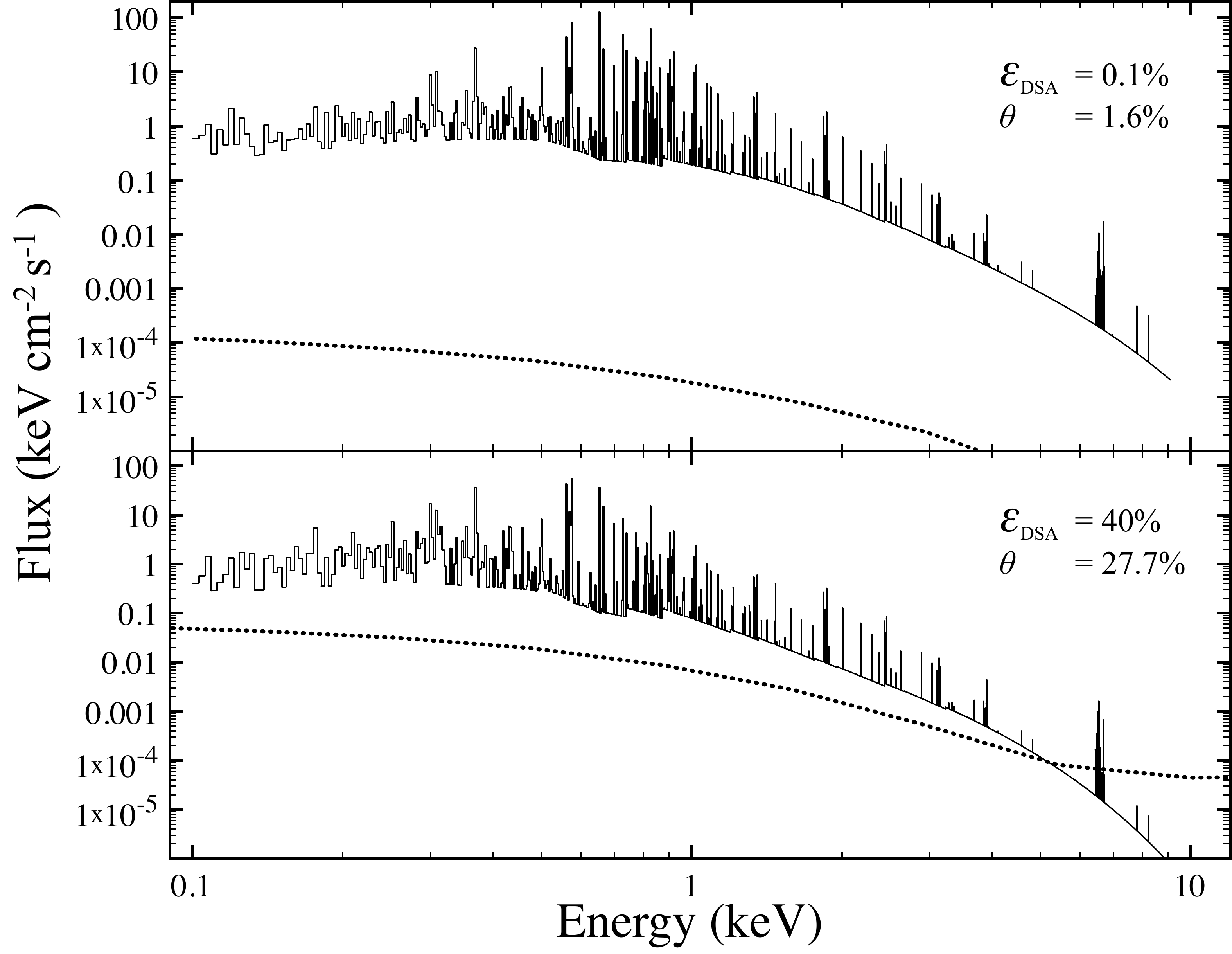}
\caption{X-ray spectral energy distribution from the TP model (top panel), and efficient acceleration case with $\varepsilon_{\text{DSA}}=40\%$ (bottom panel). The SNR models shown have input parameters $t=10^4$ yr, $E_0=1.00\times10^{51}$ erg, and $n_{\text{p}}=1.0$ cm$^{-3}$. The adopted distance to the SNRs in both cases is 5 kpc. The solid line histogram represents the thermal component of the emission, and the dotted line histogram shows the predicted non-thermal emission in this energy band.}
\label{fig:xray}
\end{center}
\end{figure}

Figure \ref{fig:xray} shows the X-ray emission spectra generated from the TP example ({\it top panel}), and the efficient acceleration case ({\it bottom panel}). In both plots, the solid histogram represents the thermal emission component, and the dotted line shows the contribution from non-thermal emission in this energy range. The thermal emission spectra are obtained using the plasma emissivity code with the NEI information obtained from CR-hydro for each example. Solar abundances are adopted for C, N, O, Ne, Mg, Si, S, Ar, Ca, and Fe, and the helium number density is fixed at 9.77\% the proton number density $n_{\text{p}}$. 

The thermal component of the spectrum in the TP case clearly dominates over the non-thermal emission in the X-ray band. The total X-ray luminosity (in the 0.3-10 keV band) is $L_{\text{X}}=2.0\times 10^{37} \text{erg } \text{s}^{-1}$, and the non-thermal contribution only represents $\sim 10^{-3}\%$ of this amount. Similarly, in the efficient acceleration case thermal emission is also dominant, however the contribution from the non-thermal spectrum is much more significant that in the TP example, and at $E>5$ keV it overtakes the thermal component, apart from a few strong emission lines. The X-ray luminosity in this case is $L_{\text{X}}= 1.2 \times 10^{37} \text{erg } \text{s}^{-1}$, and the non-thermal emission represents a 0.9\%.

\begin{figure}[h!]
\begin{center}
\includegraphics[width=\columnwidth]{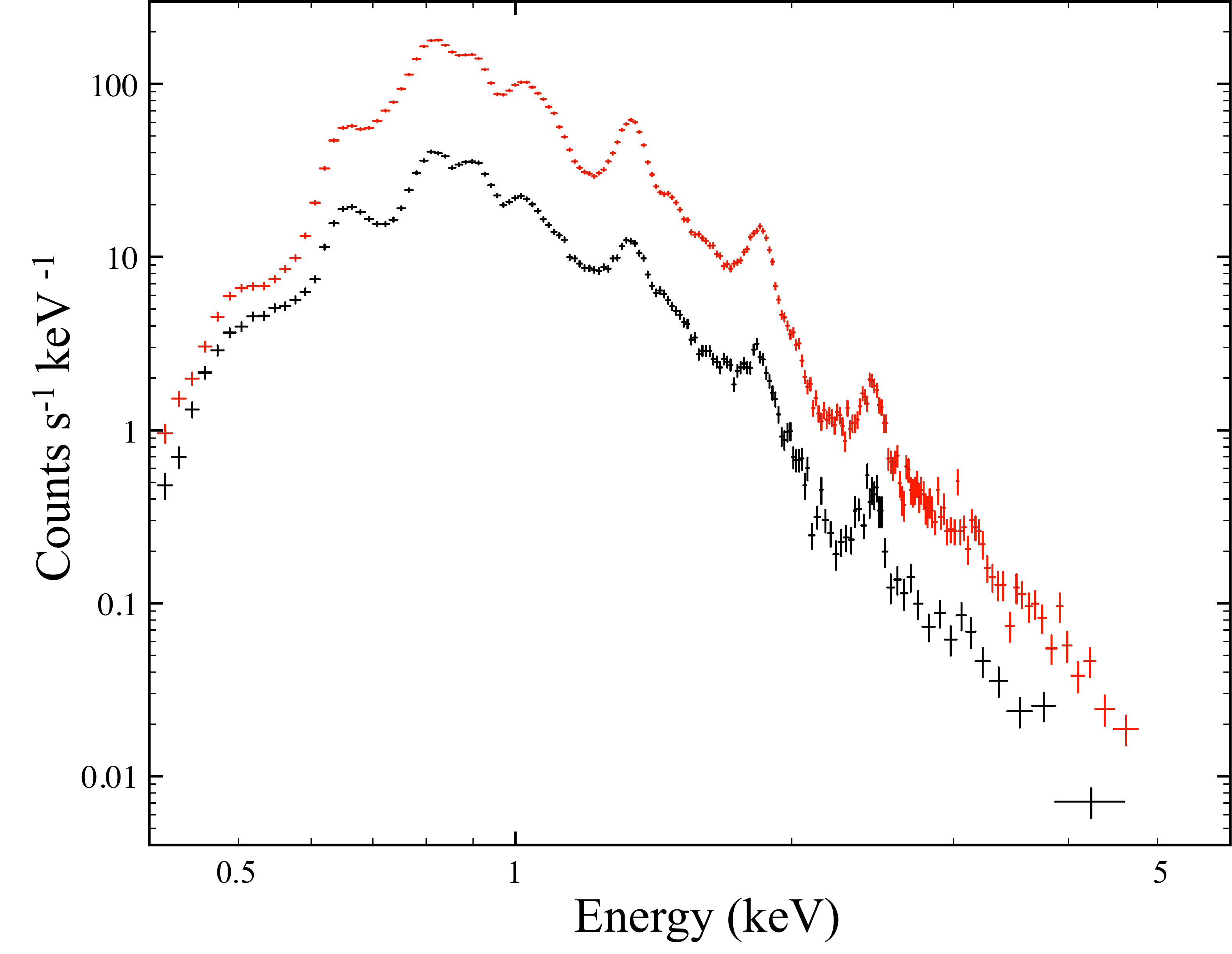}
\caption{Simulated \emph{Chandra} ACIS X-ray spectra of the model SNRs, with integrated 5 ks observation time. A Galactic column density $N_{\text{H}}=5\times10^{21}$ cm$^{-2}$ is used, and the adopted distance is 5 kpc. The SNR models shown have input parameters $t=10,000$ yr, $E_0=10^{51}$ erg, and $n_{\text{p}}=1.0$ cm$^{-3}$.The red crosses represent the TP case ($\varepsilon_{\text{DSA}}=0.1\%$) and the spectrum of the case  with efficient DSA ($\varepsilon_{\text{DSA}}=40\%$) is shown as black crosses.}
\label{fig:xray2}
\end{center}
\end{figure}

In Figure \ref{fig:xray2}, we show the X-ray emission models folded through the \emph{Chandra} ACIS instrument response. The simulated X-ray spectrum, for a 5 ks observation of the entire SNR, from the TP case is shown as red crosses, and the efficient acceleration case is presented as black crosses. Both spectra are clearly dominated by the thermal emission, and emission lines are very apparent. However, the emission from the TP case is obviously more intense than that from the case with efficient acceleration.

\subsection{Interpretation of the efficient acceleration case}

To investigate the scenario in which the emission from
an SNR with efficient particle acceleration is treated with a Sedov
analysis that neglects the effects of acceleration, we searched for
a TP case that fits the spectrum (as well as the radius) obtained
in the efficient acceleration case by investigating variations in
the input parameters such as age and SN explosion energy. The resulting X-ray models were then compared to the simulated \emph{Chandra} spectrum resulting from the efficient acceleration case detailed in Section 4.1.

An example of a TP case with results similar to those of the efficient acceleration case, is the model for a SNR with input parameters: $\varepsilon=0.1\%$; age $t=12,600$ yr; $E_0=0.56\times10^{51}$ erg; and ambient proton number density $n_{\text{p}}=1.1$ cm$^{-3}$. The resulting fraction of the explosion energy placed into CRs is $\theta=1.8\%$, and the radius of the FS is $R_{\text{FS}}=12.1$ pc.

\begin{figure}[h!]
\begin{center}
\includegraphics[width=\columnwidth]{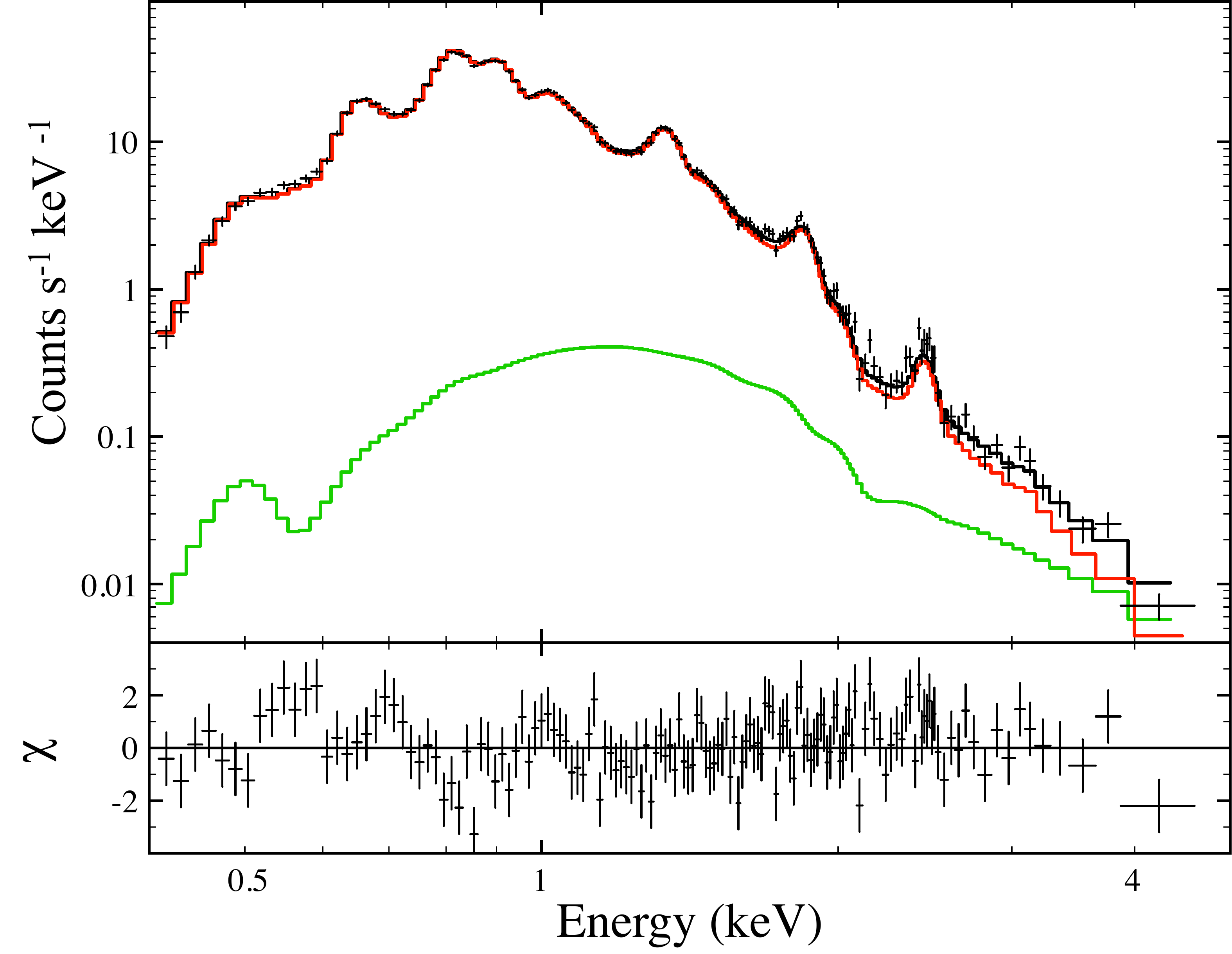}
\caption{Simulated \emph{Chandra} ACIS X-ray spectrum from the efficient acceleration case (black crosses), with integrated 5 ks observation time, with input parameters as in Figures \ref{fig:xray} and \ref{fig:xray2}. A Galactic column density $N_{\text{H}}=5\times10^{21}$ cm$^{-2}$ is used, and the adopted distance is 5 kpc. The best-fit model (black histogram), is the absorbed ($N_{\text{H}}=5.3\times10^{21}$ cm$^{-2}$) combination of a powerlaw component (green histogram), with the emission model for an SNR model with input parameters $t=12,600$ yr, $E_0=0.56\times10^{51}$ erg, and $n_{\text{p}}=1.1$ cm$^{-3}$ (red histogram). The bottom panel shows the statistical variations of the data in reference to the model.}
\label{fig:xray3}
\end{center}
\end{figure}

In Figure \ref{fig:xray3}, we show the model spectrum (red histogram) from this TP example, together with the simulated X-ray observation of the efficient acceleration case, for a 5 ks exposure (black crosses). The best-fit histogram (shown in black) is obtained assuming the model to be an absorbed combination of the thermal emission model from the simulation and a small powerlaw component (green histogram). The resulting best-fit model is obtained with parameters $N_{\text{H}}=5.3\times10^{21}$ cm$^{-2}$, powerlaw index $\Gamma=3.89$, and unabsorbed powerlaw flux (in the 0.3-10 keV band) $F_{\text{X}}= 4 \times 10^{-11} \text{erg cm}^{-2} \text{s}^{-1}$. The abundances of N (1.4 relative to solar), and S (1.2 relative to solar) were modified slightly to improve the fit. The reduced $\chi^2$ statistic of the model is 1.27 for 147 degrees of freedom, and the statistical variations of the data from the model are shown in the lower panel in Figure \ref{fig:xray3}. The TP fit appears quite reasonable, and this example illustrates that spectra from evolved SNRs where efficient CR acceleration took place can be well modeled by TP emission models, but the parameters derived from such fit would differ from the correct values. In this case the derived explosion energy is lower, and the derived density and age are higher than the input parameters, which is consistent with the trends illustrated in Figure \ref{fig:ratios2}.

\begin{figure}[h!]
\begin{center}
\includegraphics[width=\columnwidth]{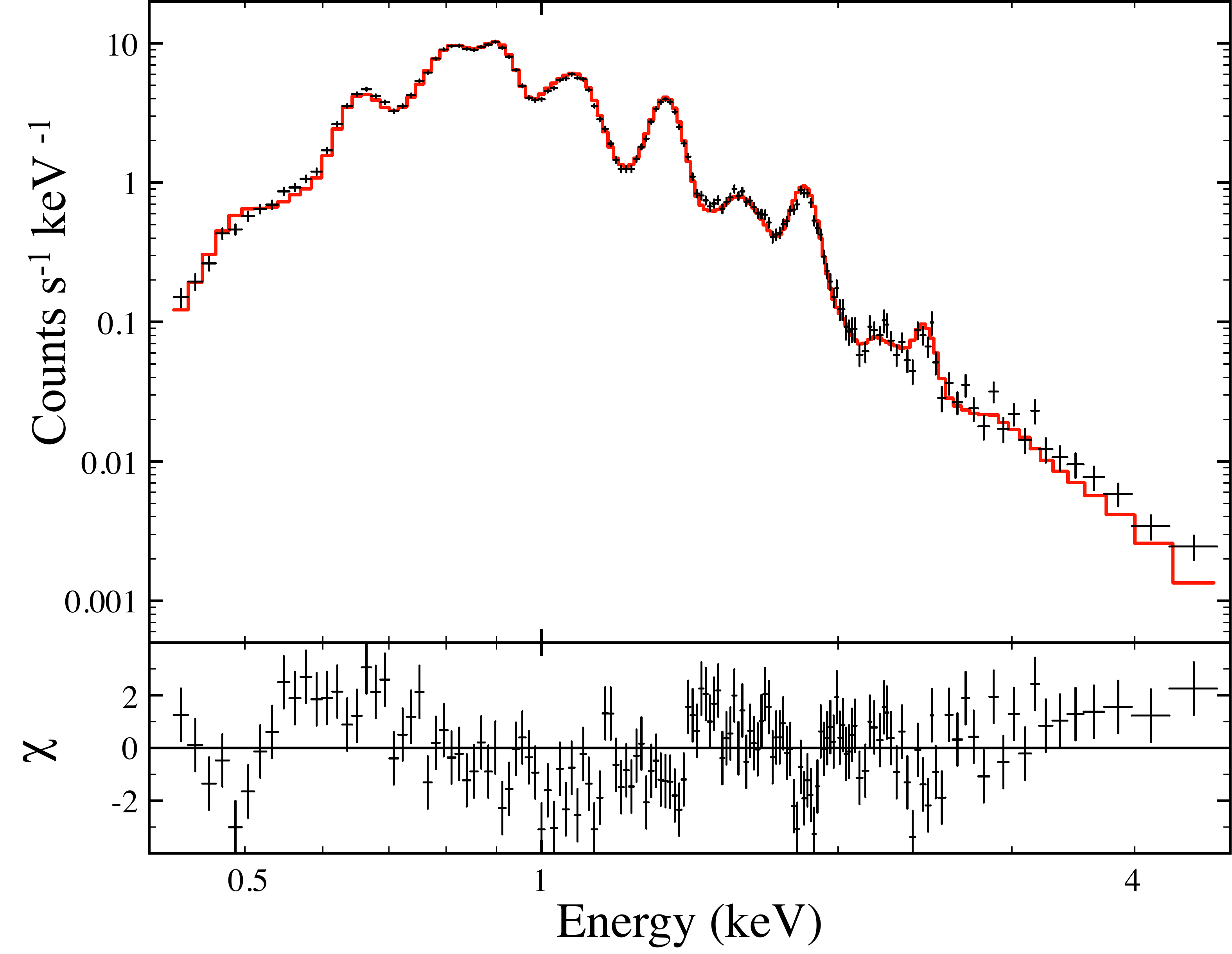}
\caption{Simulated \emph{Chandra} ACIS X-ray spectrum from an efficient acceleration case (black crosses), with integrated 20 ks observation time,  with model input parameters $t=10,000$ yr, $E_0=1.00\times10^{51}$ erg, and $n_{\text{p}}=0.1$ cm$^{-3}$. A Galactic column density $N_{\text{H}}=5\times10^{21}$ cm$^{-2}$ is used, and the adopted distance is 5 kpc. The best-fit model (black histogram), is the absorbed ($N_{\text{H}}=5.1\times10^{21}$ cm$^{-2}$) emission model for an SNR model with input parameters $t=12,200$ yr, $E_0=0.54\times10^{51}$ erg, and $n_{\text{p}}=0.1$ cm$^{-3}$ (red histogram). The bottom panel shows the statistical variations of the data in reference to the model.}
\label{fig:xray4}
\end{center}
\end{figure}

With the purpose of expanding the parameter space studied, this analysis was repeated for a similar set of initial conditions, but considering expansion into an environment with lower proton density, $n_{\text{p}}=0.1$ cm$^{-3}$. At 10,000 years, the forward shock of the DSA efficient case ($\epsilon=40\%$) expands to $R_{\text{FS}}=19$ pc, and the fraction of the total explosion energy deposited into cosmic rays is $\theta=68\%$. We find that the X-ray emission from this model is satisfactorily fit by that of a test particle case, like in the models considered above. The TP model input parameters required to reproduce the DSA efficient case are $\varepsilon=0.1\%$; age $t=12,000$ yr; $E_0=0.54\times10^{51}$ erg; and the same ambient proton number density $n_{\text{p}}=0.1$ cm$^{-3}$. This model also expands to $R_{\text{FS}}=19$ pc, and in Figure \ref{fig:xray4}, we show the model spectrum (red histogram) from this TP example, together with the simulated X-ray observation of the efficient acceleration case, for a 20 ks exposure.  The resulting best-fit model is obtained with parameters $N_{\text{H}}=5.1\times10^{21}$ cm$^{-2}$, and no powerlaw component. The abundances of N and S were varied slightly relative to solar values, as in the higher density case. The reduced $\chi^2$ statistic of the model is 2.1 for 148 degrees of freedom, and the statistical variations of the data from the model are shown in the lower panel in Figure \ref{fig:xray4}. 

%These results are reminiscent of the low SN explosion energies (below than the canonical value of $10^{51}$ erg) that been estimated from X-ray observations for several SNRs, e.g., 3C 391 \citep{Chen2001}, G349.7+0.2 \citep{Slane2002}, and G299.2-2.9 \citep{Park2007} . Our study suggests that these estimates, inferred from Sedov-Taylor analysis of remnants, could be the result of the modification of the SNR shock due to particle acceleration, and not underenergetic explosions.
These results are reminiscent of the low SN explosion energies (below than the canonical value of $10^{51}$ erg) that have been estimated from X-ray observations for several SNRs, e.g., G272.2-3.2 \citep{Harrus2001}, and G299.2-2.9 \citep{Park2007} . Our study suggests that these estimates, inferred from Sedov-Taylor analysis of remnants, could be the result of the modification of the SNR shock due to particle acceleration, and not underenergetic explosions.

\begin{figure}[h!]
\begin{center}
\includegraphics[width=\columnwidth]{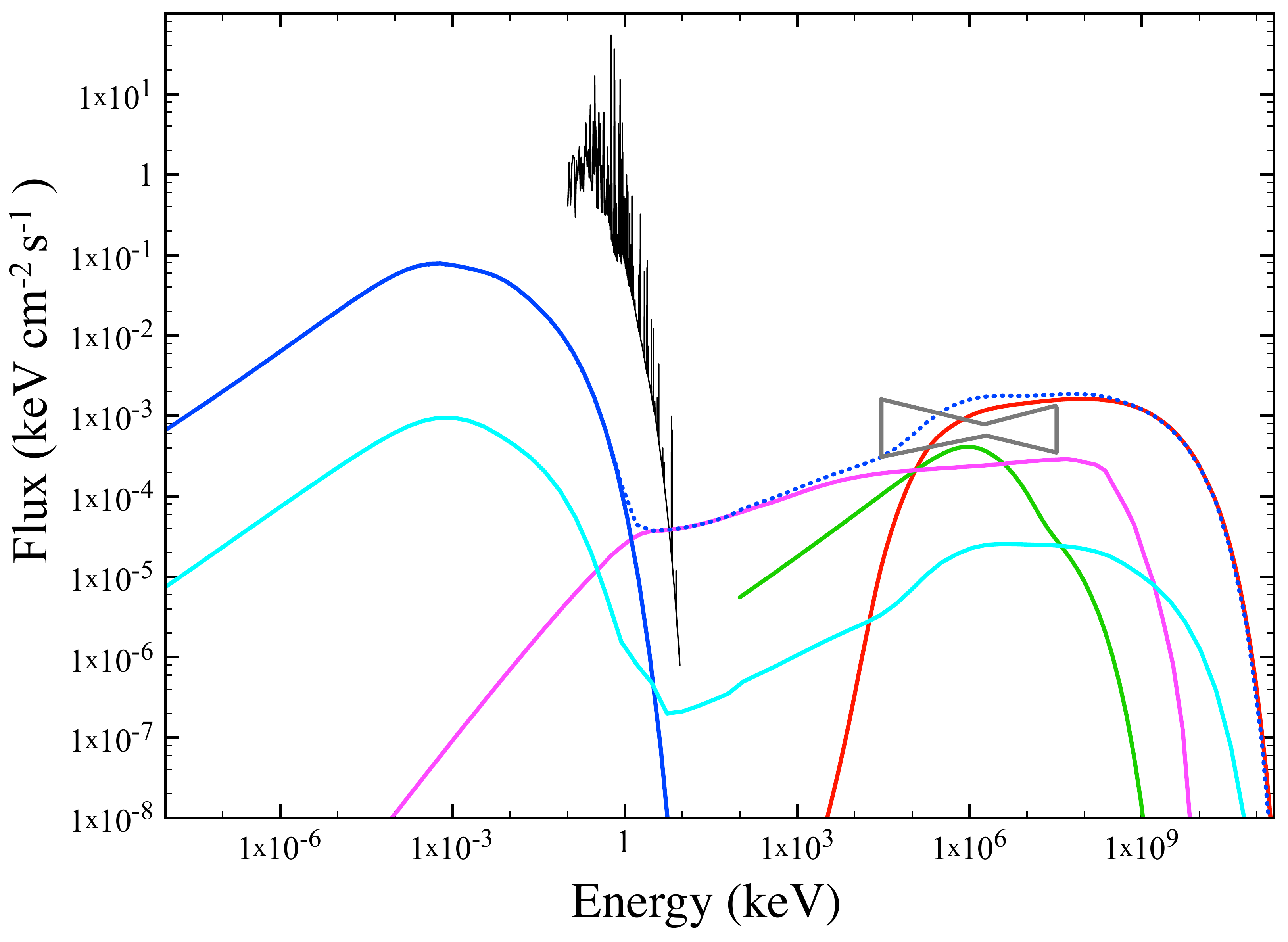}
\caption{Broadband spectrum from the model with $\varepsilon_{\text{DSA}}=40\%$, $n_0=1$ cm$^{-3}$, $E_{\text{SN}}=10^{51}$ erg, and $t=$10,000 yr. The black histogram indicates the thermal X-ray emission, and the blue solid line represents the synchrotron component. Inverse Compton radiation is shown in green, pion decay in red, and the magenta curve represents the non-thermal bremsstrahlung emission. The dotted blue line is the sum of the non-thermal emission components for the DSA efficient case, and the cyan curve shows the sum of the non-thermal radiation from the TP case, where $n_0=1.1$ cm$^{-3}$, $E_{\text{SN}}=0.56\times10^{51}$ erg, and $t=$10,000 yr. The grey bowtie indicates the integrated sensitivity of the \emph{Fermi}--LAT after 1 year. }
\label{fig:broadband}
\end{center}
\end{figure}

The broadband spectrum of the efficient acceleration example, with density $n_{\text{p}}=1$ cm$^{-3}$, is shown in Figure \ref{fig:broadband}, where we plot the thermal X-ray model, as well as each of the non-thermal emission components (synchrotron, IC, non-thermal bremsstrahlung, and pion decay), and the summed non-thermal spectrum. Additionally, the summed non-thermal emission model from the test particle case modified to fit the X-ray spectrum of the efficient DSA case is also presented. While the thermal X-ray emission from the case with efficient cosmic ray acceleration, and that of the test particle case are similar (as shown in Figure \ref{fig:xray3}), the non-thermal emission is clearly enhanced by the DSA efficiency. Figure \ref{fig:broadband} also shows, as a grey bowtie, the sensitivity after 1 year of observations of the Large Area Telescope (LAT) on board the \emph{Fermi Gamma-ray Space Telescope} \citep{Atwood2009}. Non-thermal emission from the DSA efficient case would clearly be detectable in the MeV-GeV range with the LAT, while the TP case would not bright in that energy band. 

\section{Conclusions}

We have presented a set of models of SNRs in the adiabatic phase, and considered the effects of efficient DSA on their hydrodynamics and emission characteristics. In this work we study only a limited range of parameters, yet it is clear that the production of CRs by the SNR shock significantly modifies the system. Compared to the TP case, the efficient DSA model yields smaller shock radius and speed, larger compression ratio, and lower intensity X-ray thermal emission. We also found that a model where the shock is not assumed to produce CRs can fit the X-ray observational properties of an example with efficient particle acceleration, with a different set of input parameters, and in particular a much lower explosion energy. 

The analysis of observations of thermal X-ray emission from SNRs is often done using the Sedov-Taylor interpretation, and assuming that shocks do not place a significant fraction of their energy into CRs. SN explosion energies lower than the canonical value of $10^{51}$ erg have been estimated from X-ray observations of several SNRs, including G272.2-3.2 \citep{Harrus2001}, and G299.2-2.9 \citep{Park2007} . Our study suggests that low SN explosion energies inferred from Sedov-Taylor analysis of remnants could be the result of the modification of the SNR shock due to particle acceleration.

Compared to the TP case, non-thermal emission is much more intense with efficient acceleration (see Figure \ref{fig:broadband}). Therefore, broadband modeling of SNR observations appears to be crucial for a appropriate understanding of the remnant's characteristics and those of its surroundings. 

\acknowledgements
The authors would like to thank the referee for their insightful comments. This work was carried out with partial support from Fermi Grant NNX09AT68G. P.S. and D.J.P. acknowledge support from NASA Contract NAS8-03060. D.C.E. acknowledges support through NASA grants NNH04Zss001N-LTSA and 06-ATP06-21.

\end{document}